# Distribuciones de probabilidad en las ciencias de la complejidad: una perspectiva contemporánea

Oscar Fontanelli [1,*], Pedro Miramontes [2], Ricardo Mansilla [1]

[1] Centro de Investigaciones Interdisciplinarias en Ciencias y Humanidades, Universidad Nacional Autónoma de México.

[2] Departamento de Matemáticas, Facultad de Ciencias, Universidad Nacional Autónoma de México.

[*] oscarfontanelli@ciencias.unam.mx

**Resumen**

La Ciencia en el siglo XXI está siendo dominada por nuevos enfoques que involucran a la interdisciplina, la perspectiva de sistemas y las conceptos de la teoría de la complejidad. Estos nuevos paradigmas nos obligan a dejar de lado los antiguos enfoques mecanicistas y adoptar nuevos puntos de partida basados en la aleatoriedad, la caoticidad, la estadística y la probabilidad. En este trabajo revisamos los conceptos fundamentales de la teoría de los sistemas complejos y los principales modelos probabilísticos clásicos que se utilizan en este contexto: leyes de grandes números, teorema del límite central y distribuciones normales y estables. Hablamos también de las leyes de potencias como el principal modelo para describir fenómenos con distribución de cola pesada y exploramos los principales problemas que muestran en la práctica estos modelos. Posteriormente, hablaremos de una alternativa reciente para la descripción de este tipo de fenómenos y mostraremos, por último, un par de ejemplos que ilustran el uso de este nuevo modelo.

**Abstract**

Science in the 21st century seems to be governed by novel approaches involving interdisciplinary work, systemic perspectives and complexity theory concepts. These new paradigms force us to leave aside our elder mechanistic approaches and embrace new starting points based on stochasticity, chaoticity, statistics and probability. In this work we review the fundamental ideas of complexity theory and the classic probabilistic models to study complex systems, based on the law of large numbers, central limit theorems and stable distributions. We also talk about power laws as the most common model for phenomena showing long tail distributions and we explore the principal difficulties that arise in practice with this kind of models. We show  a novel alternative for the descripition of this type of phenomena and lastly we show two examples that illustrate the applications of this new model.

## 1. Introducción

En la Ciencia del siglo XXI los enfoques de sistemas, las perspectivas interdisciplinarias y los conceptos relacionados de la complejidad parecen ir ganando terreno. Lo que comenzó en física como una forma de estudiar sistemas conformados por muchas partículas ha trascendido las fronteras de esta ciencia natural y se ha convertido en un nuevo paradigma de hacer investigación y comprender el mundo tanto en ciencias naturales como en ciencias sociales y humanidades (Mitchell, 2009).

Un concepto fundamental que se ha transformado con este cambio de paradigma es el concepto de predictibilidad. De acuerdo al enfoque clásico de Newton y Laplace, era posible, al menos en principio, predecir con exactitud la trayectoria de un cuerpo, siempre y cuando conociéramos las ecuaciones de movimiento y las condiciones iniciales. Aún así, predecir la trayectoria conjunta de más de dos cuerpos que interactúan entre sí resultó ser un asunto muy problemático, como bien apuntó en su momento el gran matemático francés Henri Poincaré.

Con el surgimiento de la mecánica estadística en el siglo XIX perdió peso este enfoque determinista para tomar un nuevo encauzamiento estadístico. Aunque las partículas microscópicas que conforman un gas siguen trayectorias newtonianas perfectamente predecibles, el sistema completo es tan grande y tan complicado que no resulta práctico resolver los sistemas de ecuaciones que los describen. Lo que sí se puede hacer es considerar las componentes microscópicas del gas como si fueran agentes aleatorios y hacer predicciones de las propiedades macroscópicas del gas como la temperatura, la presión, etc. (lo cual, por cierto, se logró llevar a cabo con enorme éxito). Tenemos así un enfoque determinista en principio, pero estadístico en la práctica.

El siglo XX fue testigo del surgimiento de una teoría física que acabó casi por completo con las ideas que se tenían sobre la predictibilidad: la mecánica cuántica. Independientemente de la interpretación que uno adopte de esta teoría (la de Copenhague, la estadística o cualquier otra), el punto es que las ecuaciones no nos permiten predecir, ni siquiera en principio, la trayectoria exacta de las partículas como electrones y fotones. Ya sea que existan elementos de incognoscibilidad intrínsecos en la naturaleza o que lo que sucede es que no conocemos aún la película completa, la mecánica cuántica nos fuerza a abandonar la esperanza de poder hacer predicciones exactas en el mundo submicroscópico de las escalas de Planck.

Aun si conociéramos el cuadro completo, es decir, las ecuaciones de movimiento y las condiciones iniciales de un sistema, la dinámica no lineal deja claro que debemos abandonar la posibilidad de predecir con exactitud, pues existen sistemas que son tan sensibles a sus condiciones iniciales que una diferencia muy pequeña entre dos valores iniciales hace que las trayectorias futuras de alejen muchísimo una de otra. Estos sistemas se llaman caóticos y nos hacen ver que, aunque las reglas que gobiernan el sistema sean deterministas, su evolución futura es estocástica para cualquier fin práctico.

Por último, en los sistemas complejos se integra lo mejor de todos los mundos: hay incertidumbre esencial en las componentes del sistemas (como ocurre en la mecánica cuántica), hay incertidumbre en las interacciones del sistema, las cuales son específicas, pueden variar con el tiempo y son, además, no lineales (con un alto grado de no linealidad). Todo parece indicar que en la Ciencia del siglo XXI debemos aceptar que no podemos predecir con exactitud el futuro, ni en la práctica ni en principio. Lo que sí podemos hacer, en cambio, es realizar predicciones probabilísticas de las propiedades macroscópicas de los sistemas complejos, y esto lo podremos hacer de mejor manera en la medida que comprendamos las propiedades estocásticas de los componentes y las interacciones de los sistemas complejos. En ello radica la enorme trascendencia de la teoría de la probabilidad en las ciencias de la complejidad.

En este trabajo haremos una descripción de los principales modelos de probabilidad que se usan en la teoría de la complejidad, discutiremos sobre sus alcances y limitaciones y hablaremos de las propuestas más recientes que existen en este campo.

## 2. ¿Qué son los sistemas complejos?

No existe una definición universalmente aceptada de qué es un sistema complejo. Sin embargo, una definición que parece ser suficientemente sencilla y general para abarcar todo lo que se suele aceptar como sistema complejo es la siguiente: lo sistemas complejos son sistemas que pueden modelarse como redes multicapas con comportamiento coevolutivo (Thurner, 2018). Aunque la variedad de sistemas complejos en ciencias y humanidades es muy extensa, algunas de las características que éstos suelen tener en común son las siguientes:

Son sistemas conformados por muchos agentes que interactúan entre sí.; estas interacciones suelen ser específicas y presentar diferentes estructuras a nivel global; las interacciones suelen depender del tiempo; son sistemas abiertos al entorno; se presentan propiedades emergentes, es decir, patrones de comportamiento global que no son totalmente explicados por las interacciones a nivel local; las dinámicas de coevolución suelen ser no lineales; son sistemas no ergódicos, es decir, los estados del sistema dependen de cómo se llegó a ellos; son sistemas altamente sensibles a sus condiciones iniciales, o sea, muestran comportamiento caótico; son sistemas que suelen presentar criticalidad autoorganizada.

No se suele pensar, por ejemplo, que un gas ideal sea un sistema complejo. Un gas ideal es un modelo en el que las partículas que conforman el gas se mueven libremente, interactuando todas con todas y dando lugar en su comportamiento colectivo a las propiedades macroscópicas del sistema, tales como la presión y la temperatura. Sin embargo, en este sistema no es necesario especificar quién interactúa con quién, pues las partículas interactúan, en principio, todas con todas y se comportan de manera idéntica. En este sentido, no decimos que la dinámica del sistema sea compleja. En cambio, una colonia de hormigas sí suele considerarse un sistema complejo, pues los agentes que lo componen (las hormigas) interactúan de manera específica, las interacciones pueden ser de varios tipos (contacto físico, intercambio de información, etcétera) y dan lugar a comportamientos colectivos que no pueden explicarse únicamente en términos de las conductas individuales: la colonia de hormigas puede construir nidos, organizarse para buscar comida, combatir con otras colonias, entre otras cosas (Solé, 1993). Son precisamente estos elementos de interacciones específicas y surgimiento de propiedades emergentes las que hacen que consideremos a la colonia de hormigas como un ejemplo clásico de sistema complejo.

Algunos ejemplos comunes de sistemas complejos son: sistemas sociales (conformados por personas), el cerebro (conformado por neuronas), los sistemas financieros (conformados, por ejemplo, por empresas), las redes metabólicas (conformadas por metabolitos), los cardúmenes (conformada por peces), etcétera. El amplio espectro de aplicaciones de la teoría de los sistemas complejos, así como su inherente carácter interdisciplinario, han despertado un gran interés en el estudio de estos sistemas. Esto, aunado al crecimiento en la capacidad de cómputo y la disponibilidad de cantidades masivas de datos, han llevado a un auge en el estudio de las ciencias de la complejidad.

Una propiedad central de los sistemas complejos es la presencia de componentes estocásticas; específicamente, hay elementos de aleatoriedad en las propiedades micro y macroscópicas de estos sistemas. Ya sea que esta aleatoriedad provenga de un nivel fundamental en las interacciones entre los componentes (por ejemplo, en sistemas cuánticos), o que provenga de nuestro desconocimiento de la estructura compleja de las interacciones o de una dinámica determinista pero caótica, el caso es que conviene, para cualquier fin práctico, estudiar las propiedades de estos sistemas como si fueran variables aleatorias. En la siguiente sección hablaremos sobre los modelos probabilísticos clásicos que se utilizan en el estudio de los sistemas complejos.

## 3. Distribuciones de probabilidad

Pensemos en el siguiente experimento: vamos a salir en este momento a la calle y vamos a entrevistar a la primer persona que veamos pasar. Nos hacemos las siguientes dos preguntas:

¿Cuál es la probabilidad de que esta persona tenga en el banco, o debajo del colchón, el doble (o más) de dinero que nosotros? Evidentemente la respuesta dependerá de cuál es nuestra situación económica actual, de en qué parte del país o de la ciudad nos encontramos en este momento, etc. Pero, en general, podríamos pensar que la probabilidad no es tan baja, no debe ser tan raro encontrarnos con alguien que tiene el doble o más del doble de dinero que nosotros. Si alguien nos dijera "conocí a un señor que tiene el doble de dinero que tú", no se nos haría extraño o inverosímil.

¿Cuál es la probabilidad de que esta persona mida el doble (o más que nosotros)? Otra vez, la respuesta va a depender de cuál es nuestra estatura, de en qué parte del mundo estamos y demás, pero bajo circunstancias que podríamos llamar "normales", la probabilidad de toparnos con alguien que nos doble la estatura es cero. ¿Es posible que nos encontremos con una persona que mida más de tres metros? De acuerdo al Libro de Récords Guiness, el ser humano más alto del que se tiene registro medía 2.72 m. Pero aun sin conocer este dato, si alguien nos dijera "conocí a un señor que mide el doble que tú", tenderíamos a creer que esto no es verdad, nos parecería inverosímil.

¿Cuál es la diferencia entonces entre la distribución del dinero y la distribución de las estaturas? Reflexionando un poco sobre este experimento pensado, parecería que las estaturas de las personas tienden a estar bastante aglutinadas alrededor de un valor promedio y que las desviaciones, hacia arriba y hacia abajo, si bien pueden llegar a ser considerables, nunca son tan grandes. Por otra parte, la cantidad de dinero que posee cada persona es un número que puede variar en un rango de escalas mucho mayor, desde cero pesos hasta los cientos de miles de millones de dólares (según la lista de los hombres más ricos de Forbes, en 2019 el fundador de Amazon, Jeff Bezos, posee una fortuna de aproximadamente 112 mil millones de dólares). En el caso del dinero, las desviaciones respecto al promedio pueden ser enormes y no es tan inusual encontrar casos extremos, o muy extremos.

Estos dos casos, la distribución de la estatura y la del dinero, son ejemplos arquetípicos de los dos grandes tipos de distribuciones que aparecen en ciencias y humanidades: las distribuciones "bien portadas", representadas por la distribución normal, y las distribuciones de colas pesadas, en las cuales hay una probabilidad no despreciable de observar eventos extremos. Vamos a hablar con un poco más de profundidad sobre estas dos clases de distribuciones.

### 3.1 Distribución normal o Gaussiana

Recordemos que una variable aleatoria $X$ es un resultado numérico que se asigna al resultado de un experimento aleatorio, es decir, un experimento cuyo resultado es incierto a priori. Una distribución de probabilidad es una función que asigna a cada posible valor de *X* una probabilidad de ocurrir. Existe una gran variedad de distribuciones de probabilidad que surgen en todas las áreas de las ciencias exactas, las ciencias sociales y las humanidades, pero de entre todas ellas la que quizá ocurre con mayor frecuencia es la distribución normal, también llamada distribución Gaussiana, en honor al gran matemático alemán Carl Friedrich Gauss.

La distribución normal queda definida a través de la siguiente fórmula:

$$f(x)=\frac{1}{\sqrt{2\pi\sigma^2}}e^{\frac{-(x-\mu)^2}{2\sigma^2}}.$$

Aquí, *f(x)* es lo que se conoce como una función de densidad de probabilidad y nos sirve para calcular la probabilidad de que la variable *X* tome un valor entre los números *a* y *b*, lo cual se hace integrando la función *f(x)* entre los límites *a* y *b*.

En esta definición, $\mu$ es un parámetro conocido como la media (el promedio) de la distribución y $\sigma^2$ es un parámetro conocido como la varianza de la distribución. Esta distribución aparece, de manera exacta y de manera aproximada, en una enorme cantidad de fenómenos en áreas tan distintas como la física (la posición de partículas tras un proceso de difusión), la medicina (la presión arterial), la biología (longitud de apéndices como pelo, garras dientes, etc.), la estadística en general (la distribución de los errores de medición en un experimento), etcétera.

La distribución normal, conocida comúnmente también como campana de Gauss, se caracteriza por ser simétrica respecto al centro (el promedio), de modo que las desviaciones hacia ambos lados son iguales, y porque decae rápidamente en ambas direcciones. Esto último implica que la probabilidad de observar valores muy, muy grandes o muy, muy pequeños (con respecto al promedio) es muy, muy baja. Entonces, si efectivamente las estaturas de una población siguen una distribución de este tipo, la probabilidad de encontrarnos con una persona cuya estatura sea el doble del promedio es esencialmente cero.

El motivo que explica la ubicuidad de la distribución normal es un teorema muy importante en teoría de probabilidad, llamado el teorema del límite central. Este teorema dice que si $X_1, \ldots X_n \ldots$ es una colección de variables aleatorias independientes, cuyo número crece a infinito, y de varianza finita (tocaremos este punto un poco más adelante), entonces la suma normalizada de todas ellas (es decir, su promedio) converge a una variable aleatoria con distribución normal.

Lo que resulta asombroso del teorema del límite central es que no dice nada acerca de la distribución que deben seguir las variables aleatorias que se están sumando (o promediando): mientras éstas tengan varianza finita, da igual qué distribución sigan, la distribución del promedio de la muestra se va a parecer cada vez más y más a la distribución normal (y será exactamente la distribución normal cuando la muestra sea infinita).

Para ilustrar cómo es que este teorema hace que la distribución Gaussiana surja en la práctica pensemos en el siguiente ejemplo: un profesor va a dar un curso de matemáticas a un grupo idealmente muy grande de estudiantes. El profesor va a evaluar el curso con los instrumentos usuales: tareas, exámenes, proyectos, participación en clase, etc. Al final, cada estudiante recibirá una calificación entre, por decir algo, 5.0 y 10.0. Aun antes de iniciar el curso, nuestro profesor puede predecir, con base en su experiencia previa, que al final del curso las calificaciones de los estudiantes seguirán aproximadamente una distribución normal, con la mayoría de las notas cercanas al promedio del grupo, algunos cuantos estudiantes sobresalientes que obtendrán 10 de calificación y algunos cuantos que reprobarán. El profesor ha observado que esto ha ocurrido en casi todos los cursos que ha dado y, en efecto, con este nuevo grupo vuelve a ocurrir. ¿Por qué las calificaciones al final siguen siempre este patrón? ¿Acaso los estudiantes se ponen de acuerdo para la mayoría sacar una nota cercana la promedio, algunos pocos sacar 10 y algunos pocos reprobar? ¿No es verdad que, al menos en principio, cada estudiante pretende sacar 10 y lo va a intentar dentro de sus posibilidades?

Lo que sucede es que la calificación final de cada estudiante depende de muchos factores: su conocimiento previo de los contenidos de la clase, su dedicación en el curso, qué tan buena es su alimentación, de qué tan lejos viene (puede ser que venga de muy lejos y llegue ya cansado a la clase), la cantidad y calidad de sus horas de sueño, su situación socioeconómica, su facilidad natural para la asignatura, etcétera. Todos estos factores pueden considerarse como variables aleatorias y todas ellas contribuyen al desempeño del estudiante, de manera que la suma o el promedio de todas ellas se traduce en la calificación final. Lo que dice el teorema del límite central es que el resultado final, la calificación en este caso, será bien aproximado por la distribución normal. Tenemos aquí un ejemplo clásico de un fenómeno colectivo, es decir, un fenómeno en el que los muchos elementos de un sistema cooperan para producir un comportamiento global que sigue ciertos patrones macroscópicos. Es precisamente este tipo de fenómenos los que presentan los sistemas complejos, lo cual explica la relevancia del teorema del límite central y de la distribución normal en las ciencias de la complejidad.

Sin embargo, no todo en la vida sigue una distribución normal, a veces ni siquiera de manera aproximada. Además de otras distribuciones "bien portadas", existe otro gran grupo de distribuciones con un comportamiento muy distinto: las llamadas leyes de Lévy, de las cuales hablaremos a continuación.

### 3.2 Distribuciones estables – Leyes de Lévy

Mencionamos que una condición necesaria para que se satisfaga el teorema del límite central es que los efectos que se están sumando, o promediando, tengan varianza finita, pero si éste no es el caso, no vamos a ver surgir la distribución normal. Pero, ¿qué significa, para empezar, que una variable aleatoria no tenga varianza finita?

Recordemos que la varianza de una variable aleatoria $X$ es el promedio de las desviaciones (cuadráticas) respecto al promedio. Si $E(X)$ es el promedio de la variable $X$, entonces la varianza está definida como $E(X) = E[(X-E(X))^2]$. Geométricamente, la varianza es una medida de qué tan ancha es la distribución. Una distribución con varianza infinita es, dicho de algún modo, "infinitamente ancha", lo cual significa que decae muy, muy lento, tan lento que la probabilidad de observar eventos extremos es relativamente alta, sin importar qué tan extremos los consideremos. En una distribución de este tipo, no es tan raro realizar observaciones que se alejan mucho del promedio, tanto así que las desviaciones cuadráticas respecto al mismo tienden a ser, en promedio, infinitas.

Para considerar el caso en que los efectos que queremos promediar puedan tener varianza infinita debemos recurrir a la versión generalizada del teorema del límite central: la suma (o el promedio) de un número, creciente a infinito, de variables aleatorias converge a una distribución de entre una familia de distribuciones llamadas leyes de Lévy (llamadas así en honor al gran matemático francés Paul Lévy). Así, el teorema del límite central de la sección anterior es un caso particular del teorema del límite central generalizado, específicamente, el caso de varianza finita.

Las distribuciones de probabilidad que resultan límites de sumas infinitas de variables aleatorias, es decir, la distribución normal y las leyes de Lévy, reciben también el nombre de distribuciones estables, pues cumplen con la siguiente propiedad: la suma de dos de ellas es una tercer variable aleatoria con la misma distribución que las originales, salvo una transformación lineal. De este modo, se dice que estas familias de distribuciones son estables ante la operación de suma. En un trabajo seminal, los matemáticos rusos Andrey Kolmogórov y Boris Gnedenko probaron que las distribuciones que son límite de suma de variables aleatorias son exactamente las distribuciones estables (Gnedenko, 1968). La distribución normal es un caso particular de distribución estable: el único caso con varianza finita.

No existe una fórmula general escribir de manera analítica la función de densidad de probabilidad de las distribuciones estables. Éstas suelen definirse a través de su llamada función característica, que es la transformada de Fourier de la función de densidad. Sin embargo, sí se sabe que la función de densidad de estas leyes decae asintóticamente de la forma

$$f(x) \sim \frac{1}{x^{\alpha+1}},$$

donde $\alpha$ es un parámetro que toma valores entre 0 y 2 (Uchaikin, 2011). Que este parámetro deba ser mayor a cero es una condición para que la función de densidad sea normalizable, es decir, para que su integral sea finita y tengamos una función de densidad bien definida. Por otra parte, este parámetro debe ser menor a dos porque, si fuera mayor o igual a dos, entonces la varianza sería finita y la suma de distribuciones de este tipo convergería, por el teorema del límite central, a una normal, así que esta distribución no sería estable. Dos casos particulares para los que sí es posible escribir la densidad mediante una fórmula cerrada son $\alpha = 2$ (una Gaussiana), $\alpha = 1$ (distribución de Cauchy) y $\alpha = 1/2$ (distribución de Lévy, que es un caso particular de las leyes de Lévy). Con el fin de ilustrar la velocidad con que decaen estas densidades, en la figura 1 mostramos la densidad de una distribución normal (con media cero y varianza uno), una distribución de Cauchy (parámetro de localización cero y parámetro de escala uno) y una distribución de Lévy (parámetro de localización cero y parámetro de escala uno).

Este tipo de decaimiento de las distribuciones estables, como el inverso de una potencia del valor $x$, relaciona esta familia de distribuciones con las llamadas leyes de potencias, de las cuales hablaremos en la siguiente sección.

## 4. Representación rango – tamaño

Quizá la manera más común de representar gráficamente las observaciones de un experimento aleatorio es a través de un histograma. En este tipo de diagrama lo que se hace es agrupar los valores numéricos de las observaciones en intervalos de igual tamaño y hacer una gráfica de barras, cuya altura indica el número de observaciones que hay en dicho intervalo. Este tipo de gráfica nos da una idea aproximada de cómo se ve la función de densidad de la cual provienen las observaciones. En la figura 2 mostramos el histograma para los datos provenientes de los dos tipos de distribuciones que hemos discutido: por un lado, una distribución "bien portada", que en este caso son las estaturas de una muestra de aproximadamente 20,000 personas en Estados Unidos (US National Health and Nutrition Examination Survey, https://www.cdc.gov/nchs/nhanes/index.htm ); por otra parte, una distribución de cola pesada, que en este caso es la fortuna de los seres humanos más ricos del mundo, de acurdo a la famosa lista de la revista Forbes en 2018 (https://www.forbes.com/real-time-billionaires/#4371b833d788).

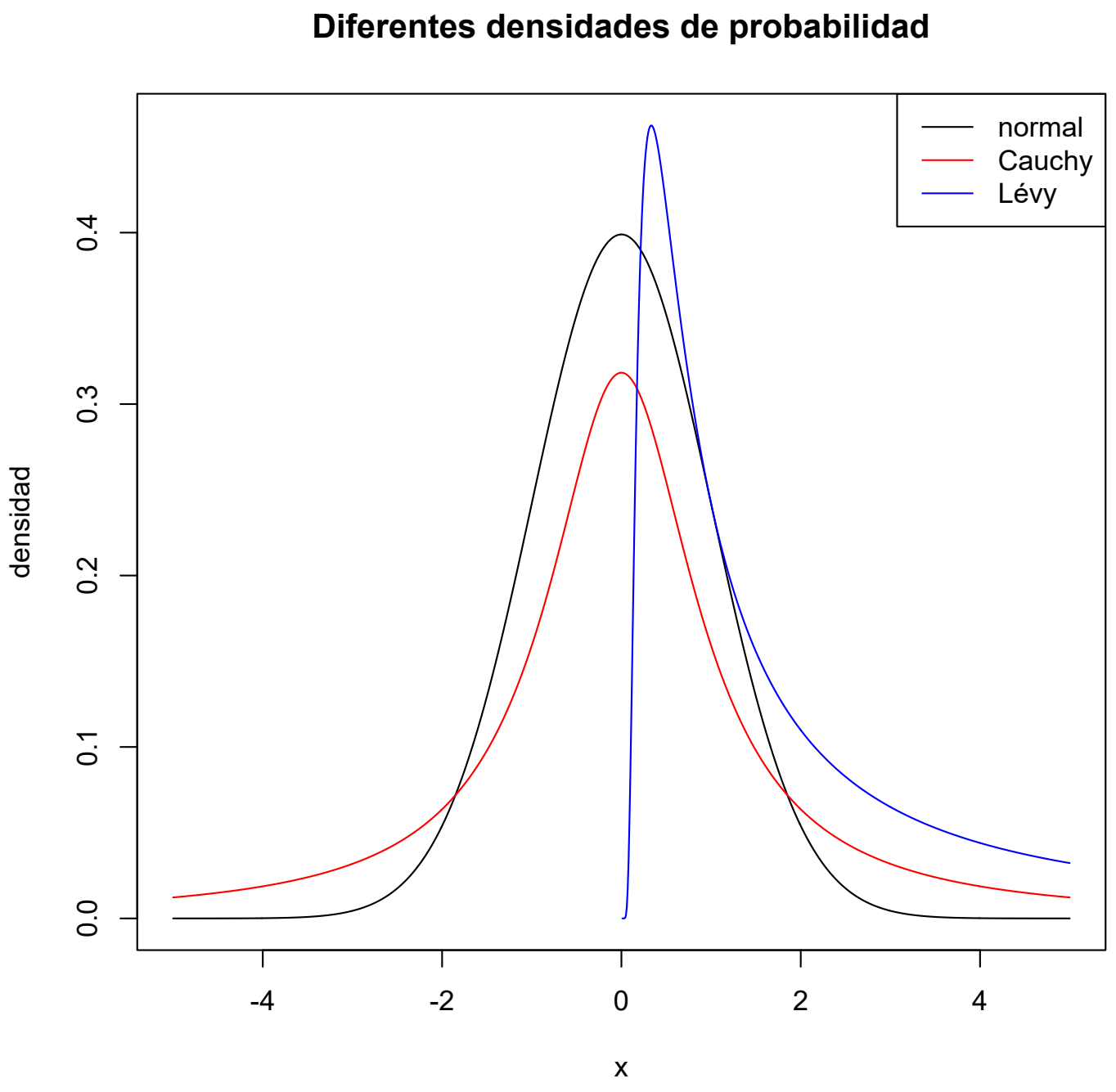

*Figura 1. Funciones de densidad de las distribuciones normal, Cauchy y Lévy.*

Nótese que, para el caso de las fortunas de los billonarios del mundo, no hemos hecho el histograma para las datos brutos de sus fortunas, sino para los logaritmos naturales de las mismas. Esto lo hemos hecho así porque, sin los logaritmos, lo que observaríamos sería una barra muy alta hasta la izquierda (correspondiente a las fortunas más grandes), tan alta respecto a las demás que el resto no se alcanzaría casi a ver (quizá el lector interesado pueda acceder a los datos brutos y hacer el histograma con el programa estadístico de su preferencia). Así, sin esta transformación, el histograma no sería una representación visual adecuada para darnos información del fenómeno. Esto sucede así porque la distribución es de cola pesada: los valores más extremos suelen ser mucho más extremos que el resto; en este caso, las fortunas más grandes son mucho más grandes que las demás.

Estos histogramas nos dan la oportunidad de seguir reflexionando sobre las diferencias entre estas dos clases de distribuciones. Pensemos que nos piden hacer una encuesta para estimar la estatura promedio y la cantidad de dinero promedio de una cierta población, para lo cual tomaremos una muestra de cien personas. Ya hemos entrevistado a 99 individuos y, justamente, el individuo 100 resulta ser el hombre más alto del mundo. Esto hará que el promedio de estatura suba, evidentemente, pero quizá sólo unos pocos centímetros; es posible que tengamos una sobreestimación de la estatura promedio, pero nada muy grave. Pero ¿qué sucedería si la persona número 100 fuera el hombre más rico del mundo? Esto sí que subiría mucho el promedio de cuánto dinero tienen estas personas, tanto que nuestra estimación ya no sería muy confiable ni muy informativa. En las distribuciones "bien portadas", el promedio y la varianza de una muestra son una buena medida del orden de magnitud de los datos, nos dan una idea de por dónde andan los valores; en cambio, en un distribución de cola pesada, ni la varianza y veces ni el promedio nos dan información sobre el comportamiento del fenómeno, pues no hay una escala característica en la que ocurra el fenómeno, y tanto el promedio como la varianza pueden crecer sin límite si la muestra es más y más grande.

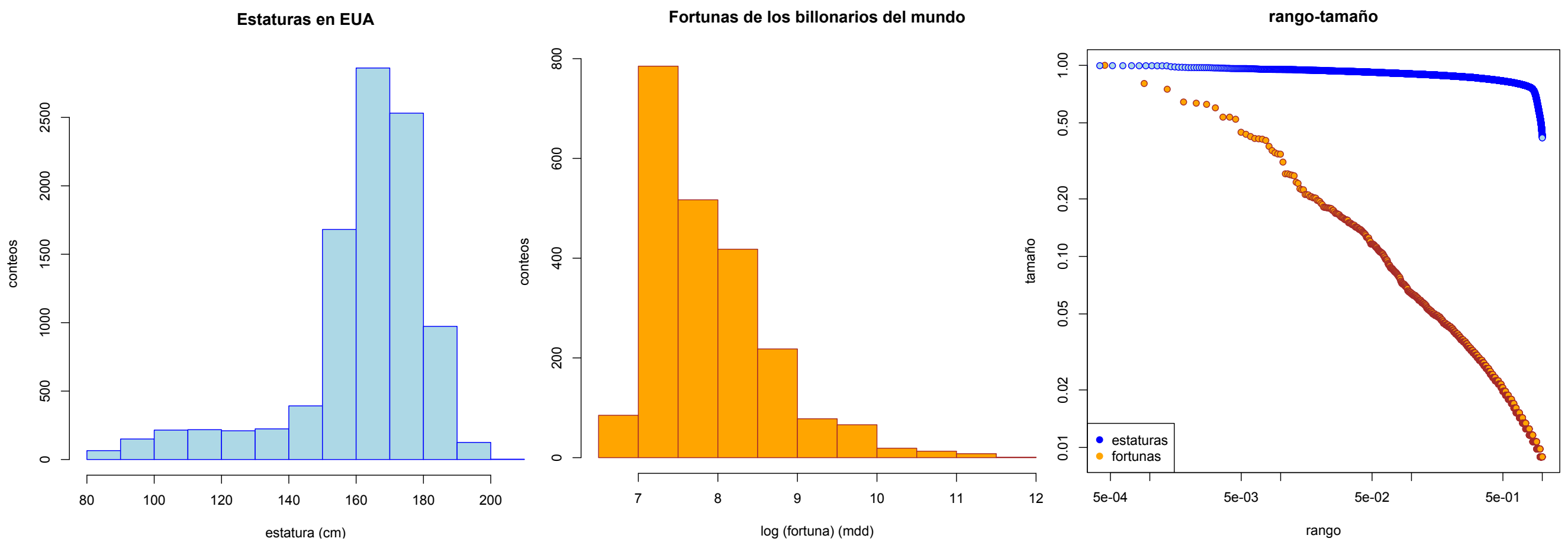

*Figura 2. Histogramas y representación rango-tamaño para la estatura de una muestra grande de pacientes en Estados Unidos y para la fortuna de las personas más ricas del mundo. Los datos de las estaturas fueron tomados de la US National Health and Nutrition Examination Survey, mientras que la información de las fortunas fue tomada de la lista de millonarios de la revista Forbes para 2018.*

Existe otra forma de representar gráficamente las observaciones de un experimento aleatorio, especialmente útil cuando éstas provienen de una distribución de cola pesada, o cuando éstas tienen una manera "natural" de ordenarse: la representación rango-tamaño.

### 4.1 La representación rango-tamaño y las leyes de potencias

Como acabamos de ver, a veces pasa que un histograma no nos proporciona una visualización adecuada de los datos que queremos analizar. Al hacer un histograma hay que elegir en cuántos intervalos vamos a dividir el rango de observaciones y, aunque hay ciertas reglas pulgar de cómo hacerlo, esta elección no deja de ser un tanto arbitraria. Una alternativa es hacer los siguiente: ordenar los datos de mayor a menor, asignar el rango 1 a la observación más grande, el rango 2 a la segunda más grande, el rango 3 a la tercera más grande y así sucesivamente, para finalmente graficar los datos ordenados como función de estos rangos. Esta gráfica se llama la *representación rango-tamaño* o *rango-frecuencia* de la muestra y una función que cuantifica la dependencia de las observaciones respecto a sus rangos se llama una *función rango-tamaño* o *función rango-frecuencia* (Sornette, 2006).

Consideremos el siguiente ejemplo, debido al lingüista estadounidense George Zipf (Zipf, 1940) (los datos que usamos aquí no son los que utilizó Zipf): la palabra que se usa con mayor frecuencia en el idioma inglés es el artículo determinado *the*; la segunda palabra más común es el verbo *be*, en cualquiera de sus conjugaciones; la tercer palabra más común es la conjunción *and.* De acuerdo al sitio *Word Frequency* (https://www.wordfrequency.info/free.asp), que analiza el Corpus de Inglés Americano Contemporáneo (cerca de 450 millones de palabras), las palabras *the, be* y *and* tienen una frecuencia de aparición aproximada de 5.0%, 2.5 % y 2.4 %, respectivamente. Asignamos los rangos 1, 2 y 3 a estas palabras, hacemos los mismo con el resto de palabras de la lista y graficamos las frecuencias relativas contra los rangos. El resultado es la gráfica que mostramos en la figura 3 (los ejes están en escala logarítmica). Tenemos aquí la representación rango-frecuencia del uso de palabras en el idioma inglés.

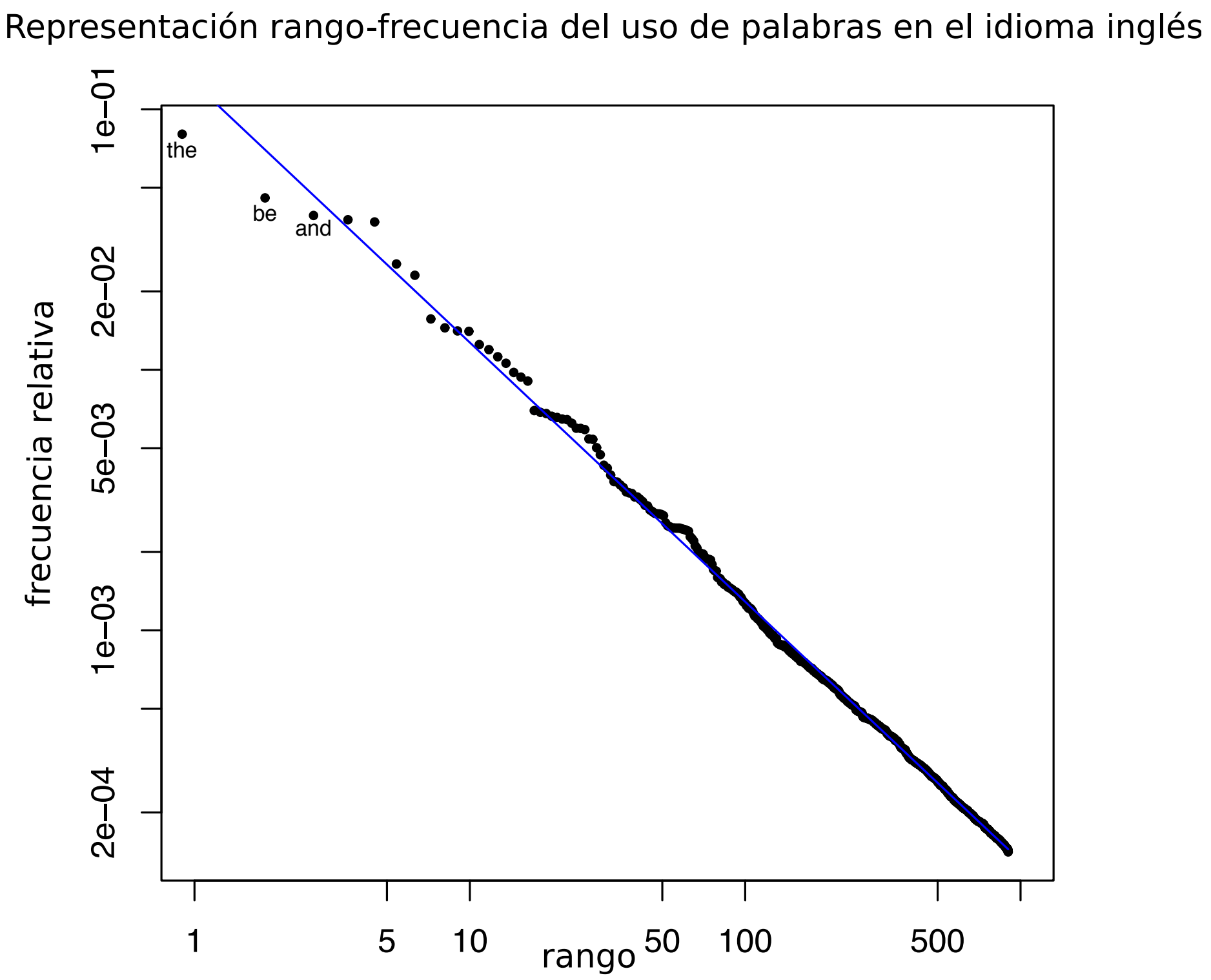

*Figura 3. Representación rango-frecuencia en escalas logarítmicas para el uso relativo de las palabras en el idioma inglés. La información se obtuvo del sitio Word Frequency, que analiza un corpus lingüístico de aproximadamente 450 millones de palabras.*

El resultado que vemos en la representación rango-frecuencia de estas palabras es notable: los puntos se ajustan sorprendentemente bien a una línea recta. Si nos detenemos a pensar un poco, nos daremos cuenta que no hay ningún motivo obvio por el cual sucede esto, los datos no tenían por qué comportarse así. Recordemos que estamos graficando con ambos ejes en escala logarítmica; que haya una aparente relación lineal entre los logaritmos de las variables (rango y frecuencia relativa) quiere decir que hay una relación entre las variables de la forma frecuencia ~ 1 / (rango)$^{\alpha}$ , donde $\alpha$ es una constante positiva. Esta relación es lo que se conoce como una ley de potencias (Schroeder, 2009).

Formalmente, una ley de potencias es una función rango-tamaño o rango-frecuencia en la cual la variable de interés varía inversamente proporcional a una potencia de su rango, es decir,

$$x(r)=\frac{k}{r^{\alpha}}.$$

Aquí $x$ es la variable que estamos observando, $r$ el rango, $\alpha$ el exponente característico y $k$ una constante de proporcionalidad. En la figura 2 vemos la representación rango-frecuencia para los ejemplos de la estatura de los estadounidenses y la fortuna de los billonarios del mundo: en ella vemos que las fortunas de los billonarios podrían ser una ley de potencias, pues los puntos parecen ajustarse razonablemente bien a una línea recta.

Las leyes de potencias tiene una propiedad interesante: son autosimilares a diferentes escalas (Newman, 2005). Esta invarianza de escala proviene del hecho de que las leyes de potencias no poseen una escala característica. Así, si un fenómeno físico es gobernado por una ley de este tipo, no tiene sentido hablar de una longitud característica o un tiempo característico. Los sistemas que siguen leyes de potencias tienen la propiedad de verse esencialmente igual a cualquier escala. Es por ello que a este tipo de fenómenos se les refiere usualmente como "libres de escala". También es interesante apuntar que los objetos fractales tiene esta propiedad de invarianza de escala; por este motivo, es muy común observar asociaciones entre leyes de potencias y geometría fractal en las ciencias de la complejidad (Mandelbrot, 1983). Sin embargo, hay que señalar que si bien una estructura fractal implica una ley de potencias, una ley de potencias no necesariamente implica una estructura fractal. Hay, en cambio, una gran cantidad de mecanismos que dan lugar a leyes de potencias, de los cuales mencionamos a continuación los más importantes: la criticalidad, los procesos de adhesión preferencial y, de nuevo, el teorema del límite central.

El término "criticalidad" se refiere a fenómenos que ocurren en puntos críticos, es decir, cuando el sistema está en una transición de fase o muy cerca de ella (Bak, 2013). Pensemos, por ejemplo, en un imán cerca de su temperatura de Curie, o sea, la temperatura por encima de la cual un material ferromagnético comienza a comportase como puramente paramagnético. Conforme vamos subiendo la temperatura, los espines siguen interactuando con sus vecinos cercanos pero comienzan a surgir cúmulos de espines alineados. Al llegar a la temperatura crítica, surgen cúmulos del tamaño del sistema completo y emergen correlaciones globales: las correlaciones entre espines, que antes sólo existían a nivel local, ahora son de largo alcance (de hecho, en la temperatura crítica las correlaciones divergen). El sistema se vuelve hipersusceptible, en el sentido de que todas las partículas perciben la magnetización global. Las propiedades físicas del sistema siguen leyes de potencias en el punto crítico; los exponentes de estas leyes de potencias se llaman los exponentes críticos del sistema. Algo que resulta notable es que la física de diferentes materiales en el punto crítico pero con exponentes críticos iguales es la misma: estos sistemas pertenecen a la misma clase de universalidad (Thurner, 2018). Otros fenómenos que siguen leyes de potencias derivadas de estados de criticalidad son los cúmulos de percolación cerca de la probabilidad de transición, las avalanchas en una pila de arena cerca del ángulo crítico y la magnitud de los terremotos.

Para hablar de los procesos de adhesión preferencial es conveniente pensar en el crecimiento de una red o grafo. Una red o grafo es una representación de un sistema de agentes (representados mediante puntos o nodos) que interactúan entre sí (las interacciones son representadas como líneas o aristas que unen los puntos). Una propiedad muy importante de los nodos de una red es su grado, que es el número de aristas adyacentes a él. Comenzamos el proceso con una red muy chiquita de muy pocos nodos, unidos entre sí de manera aleatoria. Ahora vamos a ir creciendo la red de la siguiente manera: en cada paso, añadiremos un nuevo nodo a la red pegándolo a uno de los nodos ya existentes; la probabilidad de cada nodo adyacente de atraer al nuevo nodo es proporcional a su grado. De esta forma, los nodos más conectados tienen mayor probabilidad de recibir a los nuevos nodos, con lo que quedarán aún más conectados, creando así un fenómeno tipo bola de nieve, también llamado efecto San Mateo ("porque a cualquiera que tiene, le será dado y tendrá más; pero al que no tiene, aun lo que tiene le será quitado". Mateo 13:19). Este proceso, conocido como adhesión preferencial, da lugar a una red en la cual los grados de sus nodos siguen una ley de potencias (Barabási, 1999). Algunos fenómenos que siguen una ley de potencias producto de un proceso de adhesión preferencial son las ventas de libros, los números de clicks en las páginas de Internet y la conectividad de la *world wide web.*

Finalmente, si un fenómeno aleatorio sigue una ley de potencias con exponente $\alpha$, entonces su densidad de probabilidad debe decaer de la forma

$$f(x) \sim \frac{1}{x^{\alpha+1}},$$

Vemos aquí que, si el exponente $\alpha$ está entre 0 y 2, éste es precisamente el decaimiento de una distribución estable. Y ya sabemos que las distribuciones estables surgen de manera natural en el contexto de la versión generalizada del teorema del límite central. Es por lo tanto natural esperar que si una propiedad de un sistema es el resultado de la suma o el promedio de muchas contribuciones aleatorias, algunas de ellas de varianza infinita, entonces esta propiedad será bien representada por una ley de potencias. Aquí es importante señalar que si el exponente de la ley de potencias no está entre 0 y 2, entonces este comportamiento no puede ser atribuido al teorema del límite central.

Las leyes de potencias poseen un cierto nivel de ubicuidad en las ciencias exactas, las ciencias sociales y las humanidades, además de que poseen propiedades muy interesantes y que se comprende bastante bien cómo es que éstas pueden surgir en diversos contextos. Lo que quizá no es tan sencillo es identificarlas plenamente al momento de analizar datos empíricos, lo que ha llevado a una sobreexplotación, por decirlo de alguna manera, de estos modelos. De esto hablaremos en la siguiente sección.

### 4.2 El problema con (algunas) leyes de potencias.

Al revisar la literatura, uno puede encontrar reportes de leyes de potencias en áreas que van desde la biología molecular hasta las finanzas. El procedimiento usual en la gran mayoría de estos trabajos para detectar y probar una ley de potencias en datos empíricos es el siguiente: realizar la gráfica de rango-tamaño o rango-frecuencia en escala logarítmica, ver si los puntos en esta gráfica se pegan bien a una línea recta y, en caso de que sea así, estimar el exponente mediante una regresión lineal sobre los logaritmos de las variables, usando el coeficiente de determinación de la regresión como medida de bondad de ajuste.

Hay varios problemas con este método: el primero es que las gráficas en escala logarítmica suelen ser engañosas, pues las distancias no son iguales en diferentes lugares del plano y lo que parecería ser una diferencia muy pequeña entre dos puntos, es en realidad muy grande. Esto nos lleva al segundo problema: como las distancias entre los puntos y la recta ajustada no son iguales en diferentes regiones del plano, los errores no se ponderan igual (este problema puede corregirse introduciendo alguna variante del método de mínimos cuadrados ponderados o una regresión no lineal). En tercer lugar, el coeficiente de determinación ya no es una buena medida de la bondad de ajuste, por lo mismo de que hay una subestimación muy grande de las desviaciones para valores altos del rango.

Todos estos problemas hacen que detectar y probar una ley de potencias en datos empíricos sea una tarea que no es sencilla. De hecho, cuando se han utilizado métodos estadísticos más sofisticados para estimar el coeficiente y medir la bondad de ajuste, se han rechazado muchas leyes de potencias que se habían reportado previamente y cuya existencia se daba prácticamente por sentada (Clauset, 2009). Algunos ejemplos de esto son los tamaños en kilobytes de las páginas de Internet, el grado de los nodos (metabolitos) en redes metabólicas, el número de clics en páginas de Internet y el número de hipervínculos que apuntan hacia sitios de Internet. Por otra parte, ejemplos de fenómenos donde las leyes de potencias sí son modelos estadísticos adecuados son la población en ciudades grandes, la intensidad de las manchas solares, el número de artículos de investigación por investigador y las leyes de escalamiento alométrico.

Otro problema que se observa en muchas supuestas leyes de potencias que ajustan bien datos empíricos es que éstas solo ajustan bien una régimen de la distribución, específicamente, el régimen de rangos pequeños u observaciones grandes. Es muy común observar que los datos siguen una ley de potencia en este intervalo, pero decaen el el régimen de rango altos u observaciones pequeñas. Esta falla en la cola de la distribución suele atribuirse a efectos de tamaño finito: una ley de potencias exacta sólo puede surgir cuando el sistema es de tamaño infinito, o cuando se dispone de una cantidad infinita de energía, o cuando puede haber terremotos de magnitud infinita o cuando se suma una cantidad infinita de variables aleatorias, y sucede que en la vida real las propiedades físicas de cualquier sistema son siempre cantidades finitas (Stumpf, 2012). En la figura 4 mostramos un ejemplo de este quiebre un la ley de potencias. Aquí lo que graficamos es la población de ciudades y distritos de la India. En ambos casos, pero quizá sea más claro para las ciudades, vemos una aparente línea recta al inicio que cae abruptamente pasando un cierto valor del rango.

En general, para poder afirmar que un cierto fenómeno es gobernado por una ley de potencias (o por cualquier otro modelo estadístico o probabilístico) deben cumplirse dos cosas: en primer lugar, debe haber evidencia estadística sólida en favor del modelo (lo cual, como hemos mencionado, no en sí mismo un asunto trivial); en segundo lugar, debe existir en mecanismo teórico que explique por qué la cantidad que estamos analizando debe seguir una ley de potencias. Con respecto al segundo punto, los efectos de tamaño finito que hemos mencionado son reales, todos los sistemas en la naturaleza y en las humanidades son de tamaño finito, pero entonces debe estudiarse cuidadosamente en qué dominio de escala podemos esperar una ley de potencias y en qué dominio no. Es factible que estos sistemas tengan al menos dos regímenes, cada uno dominado por dinámicas distintas, y que sea entonces necesario introducir una corrección en todo el cuerpo de la distribución (Laherre, 1998). Una alternativa que se propuso hace unos años y que se ha utilizado con cierto éxito en la descripción de numerosos fenómenos naturales y humanos es la Distribución Beta Discreta Generalizada. De esto hablaremos en la siguiente sección.

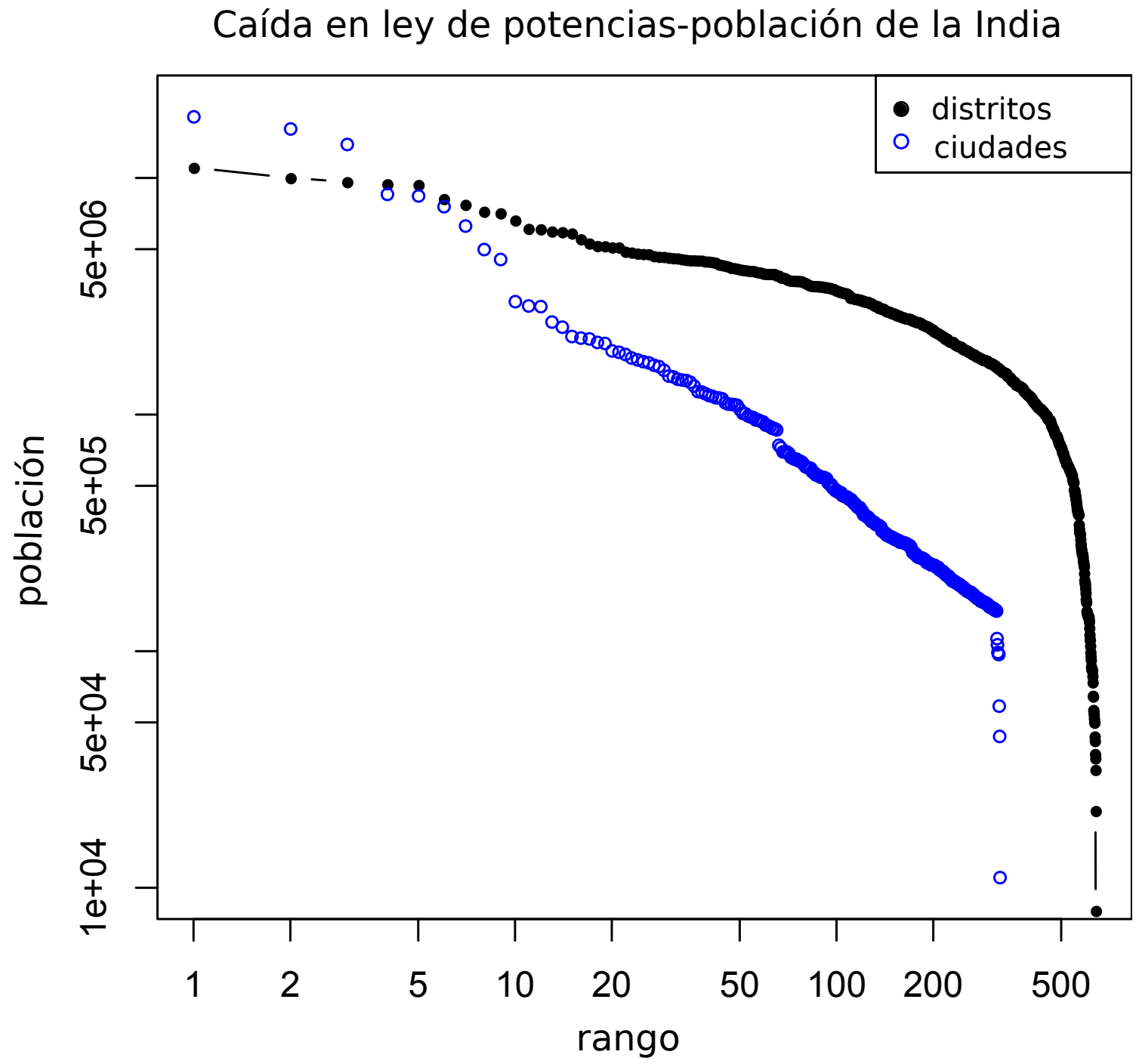

*Figura 4. Representación rango-tamaño en escalas logarítmicas de la población en distritos y ciudades de la India. Los datos fueron tomados del sitio City Population (https://www.citypopulation.de/India.html ).*

## 5. Una alternativa: la función DGBD

La Distribución Beta Discreta Generalizada es la función rango tamaño definida por

$$x(r) = A\frac{(N+1-r)^b}{r^a}.$$

Aquí $x$ es la variable de interés, $r$ el rango, $N$ el rango máximo (en ausencia de empates, $N$ es igual al número de observaciones en la muestra), $A$ es una constante de normalización, mientras que $a$ y $b$ son parámetros. Esta función fue propuesta originalmente por Germinal Cocho en 2007 para corregir las desviaciones a leyes de potencias en la distribución de los factores de impacto de revistas académicas (Mansilla, 2007). Observamos que esta función se reduce a una ley de potencias en el caso $b = 0$. El parámetro $a$ controla el régimen en el cual la DGBD se comporta como una ley de potencias (rango bajos y observaciones grandes), mientras que el parámetro $b$ controla las desviaciones en el dominio de rangos altos y observaciones pequeñas.

Como hemos mencionado, esta función se utilizó en un principio para ajustar los factores de impacto de una muestra de revistas académicas. En este trabajo, se concluyó que éste es un mejor modelo que la ley de potencias, la distribución de Mandelbrot y la función de Lavalette (esta última es un caso particular de la DGBD). En este contexto de métricas académicas, la DGBD se ha usado para describir las cantidades de citas que reciben investigadores de diferentes campos, concluyendo que existen regularidades estadísticas que pueden usarse para evaluar el progreso de carreras académicas (Petersen, 2011).

Este modelo también se ha puesto exitosamente en práctica en el campo de la lingüística. En esta área, hay estudios que que muestran que la DGBD es un mejor modelo que las leyes de potencias y que otras funciones de dos parámetros para describir la distribución del uso de palabras en las novelas *Moby Dick* y *El ingenioso hidalgo Don Quijote de la Mancha* (Li, 2010). Asimismo, es un modelo estadísticamente adecuado para describir el uso de palabras en discursos presidenciales de México y Estados Unidos durante varias décadas (Li, 2011), así como el número de caracteres por sílaba en el idioma mandarín (Li, 2012).

En lo que respecta a fenómenos sociales, el número de adeptos a las principales religiones del mundo durante el siglo XX parece seguir una DGBD (Ausloos, 2014). Se ha observado también que la población de unidades administrativas de segundo nivel (municipios, condados, etc.) sigue este modelo, el cual resulta superior a la ley de potencias y a la distribución lognormal (Fontanelli, 2017). Hablando de manifestaciones artísticas, se ha reportado que la DGBD ajusta bien la frecuencia relativa de aparición de las notas musicales en una muestra de más de de 1,800 obras de diferentes épocas y estilos (del Río, 2008). Más aún, los tamaños característicos de figuras geométricas en cuadros abstractos parecen seguir también esta función (Martínez-Mekler, 2009). En lo que respecta a fenómenos financieros, también se ha reportado que las magnitudes de las caídas de los mercados de bolsa estadounidenses siguen una DGBD (Martínez-Mekler, 2009).

Todos estos ejemplos evidencian la utilidad de esta función para ajustar datos con el comportamiento del que hablábamos: una aparente ley de potencias que se cae en el dominio de observaciones muy pequeñas. El creciente número de fenómenos que son bien descritos por este modelo y la gran diversidad de los mismos son para algunos autores ya un signo de un cierto nivel de ubicuidad de la función DGBD. No obstante, hemos subrayado que deben cumplirse al menos dos cosas para poder afirmar que un fenómeno sigue un cierto modelo: debe haber evidencia estadística sólida en favor del modelo y debe existir un mecanismo teórico que explique por qué dicho modelo es plausible. En algunos casos, parece que la DGBD pasa la primera prueba. ¿Qué podemos decir respecto al segundo criterio?

Hay varias propuestas de mecanismo generadores de la función DGBD; de hecho, la mayoría de la líneas de investigación abiertas en este tema van en este sentido. El primer mecanismo que se propuso fue un modelo de expansión-modificación que involucra los mecanismo básicos de la evolución neutral de una secuencia genómica: duplicaciones y mutaciones puntuales (Li, 1991). En este proceso se comienza con una serie de variables que pueden tomar los valores 0 y 1. En cada paso del proceso, las variables se modifican de acuerdo a la siguiente regla: con probabilidad $p$ la variable se duplica y con probabilidad $1-p$ la variable cambia su valor. Este proceso genera una cadena de ceros y unos en la cual la longitud de subcadenas de elementos consecutivos (rachas de ceros y de unos) tiene una distribución rango-tamaño bien descrita por la DGBD.

Se ha propuesto también un modelo probabilístico, el cual consiste en un proceso estocástico de resta sucesiva de variables aleatorias independientes, condicional a un valor positivo de la diferencia. Utilizando simulaciones estocásticas, los autores de este modelo encontraron evidencia numérica de que este proceso posee un atractor y que la distribución rango-tamaño del mismo se ajusta bien con la función DGBD (del Río, 2011).

Una tercer propuesta es a través de ecuaciones maestras para modelar la evolución temporal de redes complejas con fenómenos de nacimiento y muerte (Álvarez-Martínez, 2014). En este modelo, se propone una ecuación para la dinámica de una red con probabilidades de transición positivas y negativas, la cual puede aproximarse como una ecuación tipo Fokker-Planck. En este trabajo, los autores prueban que la solución estacionaria de esta ecuación puede ser bien representada por la función DGBD.

Finalmente, una cuarta propuesta, basada en las observaciones empíricas de que los datos poblaciones de unidades administrativas en países de todo el mundo son muy bien representadas por la DGBD, es el llamado modelo de "dividir y pegar" (Li, 2016). En este mecanismo, uno comienza con una distribución espacial aleatoria de unidades con un número positivo asociado a cada una de ellas (hay que imaginarlo como municipios y su número de habitantes). En cada paso del proceso, se decide de manera semialeatoria dividir algunas unidades en dos, o bien pegar dos unidades vecinas formando una más grande, emulando así los procesos de seccionar municipios muy grandes o de fusionar municipios vecinos que ocurren en todos los países del mundo. Se ha observado que este proceso posee un atractor, es decir, converge a una cierta distribución independientemente del punto de partida, y que la distribución rango-tamaño de este atractor es precisamente una DGBD (Fontanelli, 2017).

Todo lo anterior sugiere que la DGBD es una alternativa viable y digna de ser explorada en fenómenos que no siguen claramente ni una distribución "bien portada" ni una ley de potencias. Este tipo de comportamiento lo podemos detectar mediante la representación rango-tamaño de datos empíricos o experimentales, en la cual observamos una ley de potencias que se quiebra pasando un cierto punto. Para terminar este trabajo, mostraremos dos ejemplos de este tipo de fenómenos.

## 6. Ejemplos: datos poblacionales y series financieras.

Como primer ejemplo tomemos la distribución poblacional de las ciudades en Estados Unidos. Históricamente, no se tenía información sobre la población de los asentamientos más pequeños como pueblos y villas, por lo cual sólo se podían observar los datos de ciudades medianas y grandes. Lo que se observaba era una ley de potencias, pero como hemos mencionado, ocurre con mucha frecuencia que estas gráficas se doblan al llegar, en este caso, a las ciudades de menor población, sólo que esto no se veía porque no había datos suficientes. Sin embargo, censos más recientes arrojaron información sobre estos lugares de muy baja población y se abrió el debate de qué modelo describía mejor los datos: la distribución lognormal o una ley de potencias (Eeckhout 2004, Levy, 2009, Eeckhou, 2009). Ante esta dicotomía, nosotros postulamos que la DGBD puede servir como una tercer alternativa.

Analizamos datos oficiales obtenidos a través del sitio de la Oficina del Censo de los Estados Unidos (https://www.census.gov/). Usando los datos poblacionales del censo de 2010 a nivel condados (unidades administrativas artificiales) y ciudades, pueblos y villas (agregados poblacionales orgánico), mostramos el histograma y la representación rango-tamaño en escala logarítmica. Ajustamos también a cada conjunto de datos la distribución lognormal, estimando sus parámetros mediante máxima verosimilitud, y la función DGBD, estimando sus parámetros mediante una regresión no lineal con el algoritmo de Levenberg-Marquardt. En la figura 5 mostramos los resultados.

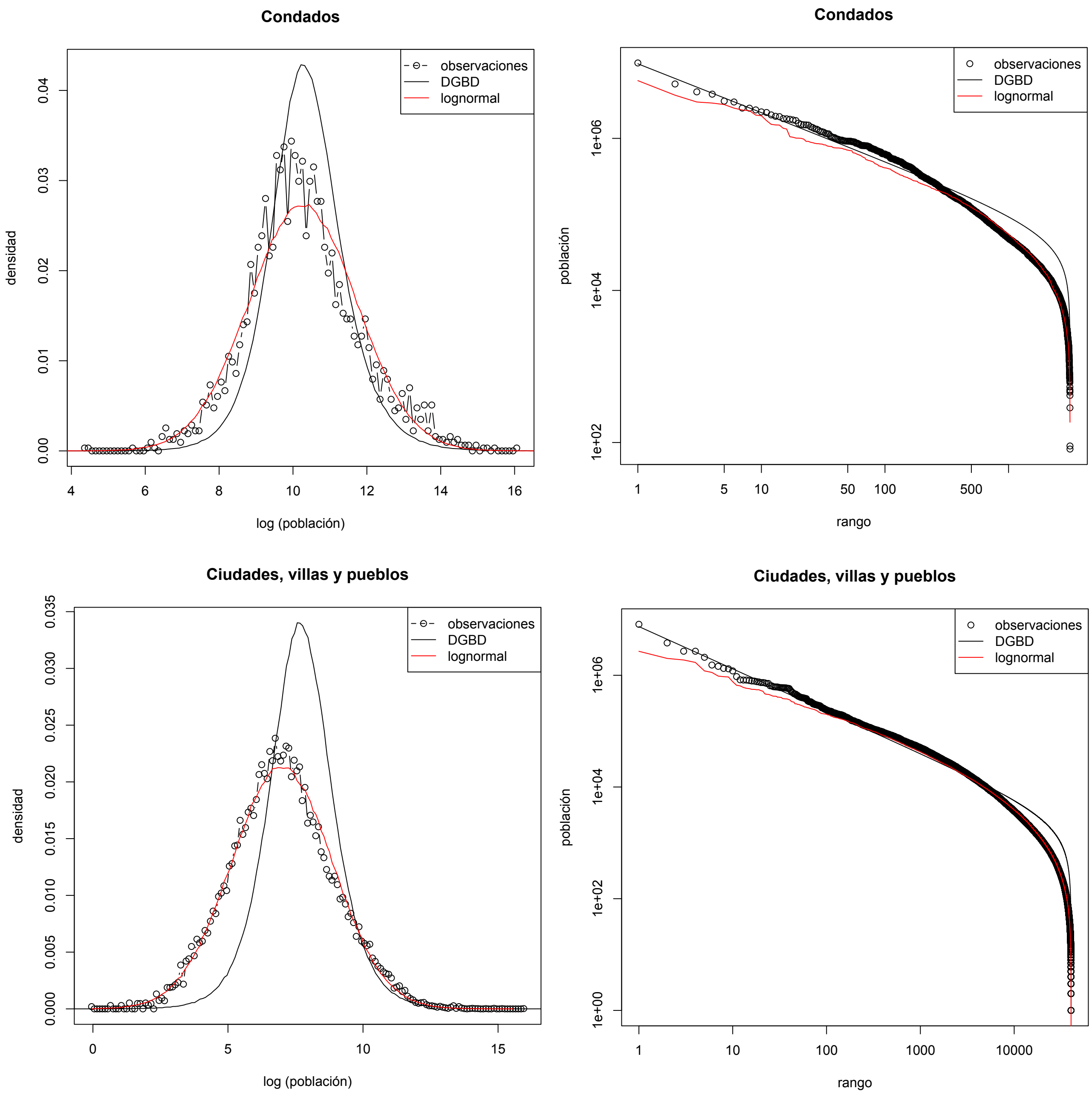

*Figura 5. Histogramas y representación rango-tamaño en escala logarítmica de la distribución poblacional en ciudades y en condados de Estados Unidos. En todos los casos mostramos los datos empíricos, así como los ajustes a la distribución lognormal y a la función DGBD.*

Un simple examen visual de las gráficas rango-tamaño parece descartar la ley de potencias. Si bien es verdad que observar una línea recta no implica necesariamente una ley de potencias, no observarla sí la descarta. En los histogramas podemos observar que la distribución lognormal ajusta mejor el cuerpo de la distribución, es decir, las ciudades o condados de tamaños relativamente pequeños, mientras que la DGBD es mejor para describir lo que pasa en la cola pesada de la distribución, lo cual puede observarse en la parte izquierda de los diagramas rango-tamaño. Finalmente, el corrimiento hacia la derecha de la densidad de la DGBD respecto a los datos podría ser indicador de que existe un sesgo en los estimadores de los parámetros, haciendo deseable el desarrollo de nuevos métodos de estimación para esta función.

Como segundo ejemplo analizamos los rendimientos del índice Nasdaq 100, el cual incluye a cien de las mayores compañías no financieras que cotizan en el mercado de valores Nasdaq. De acuerdo al modelo clásico, los rendimientos deberían seguir una distribución lognormal (y por lo tanto, los rendimientos logarítmicos una distribución normal). Pero como hemos señalado, estas distribuciones decaen muy rápido y asignan probabilidades casi nulas a los eventos extremos, como las caídas abruptas y los *crashes* financieros. Sin embargo, estos eventos ocurren y ocurren además con una frecuencia mayor que la predicha por los modelos clásicos, por lo que surge de la necesidad de modelarlos con distribuciones de cola pesada (Mantegna, 1995). Se ha visto que, en general, los eventos financieros extremos son bien descritos por leyes de potencias, así que estamos de nuevo ante un fenómeno que se comporta en un régimen como una lognormal y en otro como una ley de potencias. Nosotros creemos que es precisamente este tipo de comportamiento el que es bien representado por la DGBD.

En la figura mostramos el histograma y la representación rango-tamaño de los rendimientos financieros diarios, usando el precio más alto de cada día, para una serie de 30 años del índice Nasdaq 100. Los datos los obtuvimos a través del portal de *Yahoo Finance* (https://finance.yahoo.com/ )

La inspección visual de la rango-tamaño nos muestra una vez más el fenómeno de ley de potencias que se quiebra pasando un cierto punto. Este comportamiento parece ser universal en muchos sistemas complejos. En el histograma observamos nuevamente que la lognormal modela mejor el centro de la distribución, es decir, los rendimientos "usuales", y aquí no es distinguible qué sucede en las colas. Esta parte se ve mejor en la rango-tamaño, y lo que observamos es que ambos modelos se quedan cortos al predecir las caídas extremas (en el llamado "lunes negro" de 1987, el índice Nasdaq cayó en aproximadamente 11%, lo cual fue aun menos de lo que se precipitaron otros indicadores como el Dow Jones, que perdió 22.6%). Aún así, la DGBD se acerca un poco más a los datos en esta parte de la distribución.

## 7. Conclusiones y perspectivas a futuro.

Los actuales paradigmas en la ciencia del siglo XXI parecen dirigirnos a nuevos terrenos en los cuales los pilares fundamentales de la investigación serán los enfoques de sistemas, las perspectivas interdisciplinarias y las conceptos de las teorías de la complejidad. Una característica central de los sistemas complejos es que no podemos predecir sus propiedades macroscópicas ni su evolución temporal con exactitud, ni en la práctica ni en principio. Por lo tanto, si estos nuevos enfoques prevalecen, será menester que abdiquemos de las teorías puramente deterministas y que nos encaucemos a ideas y metodologías que involucren de manera esencial al azar y la aleatoriedad; en ello jugarán un papel trascendental la teoría de la probabilidad, la estadística y las emergentes áreas de ciencia de datos y aprendizaje profundo.

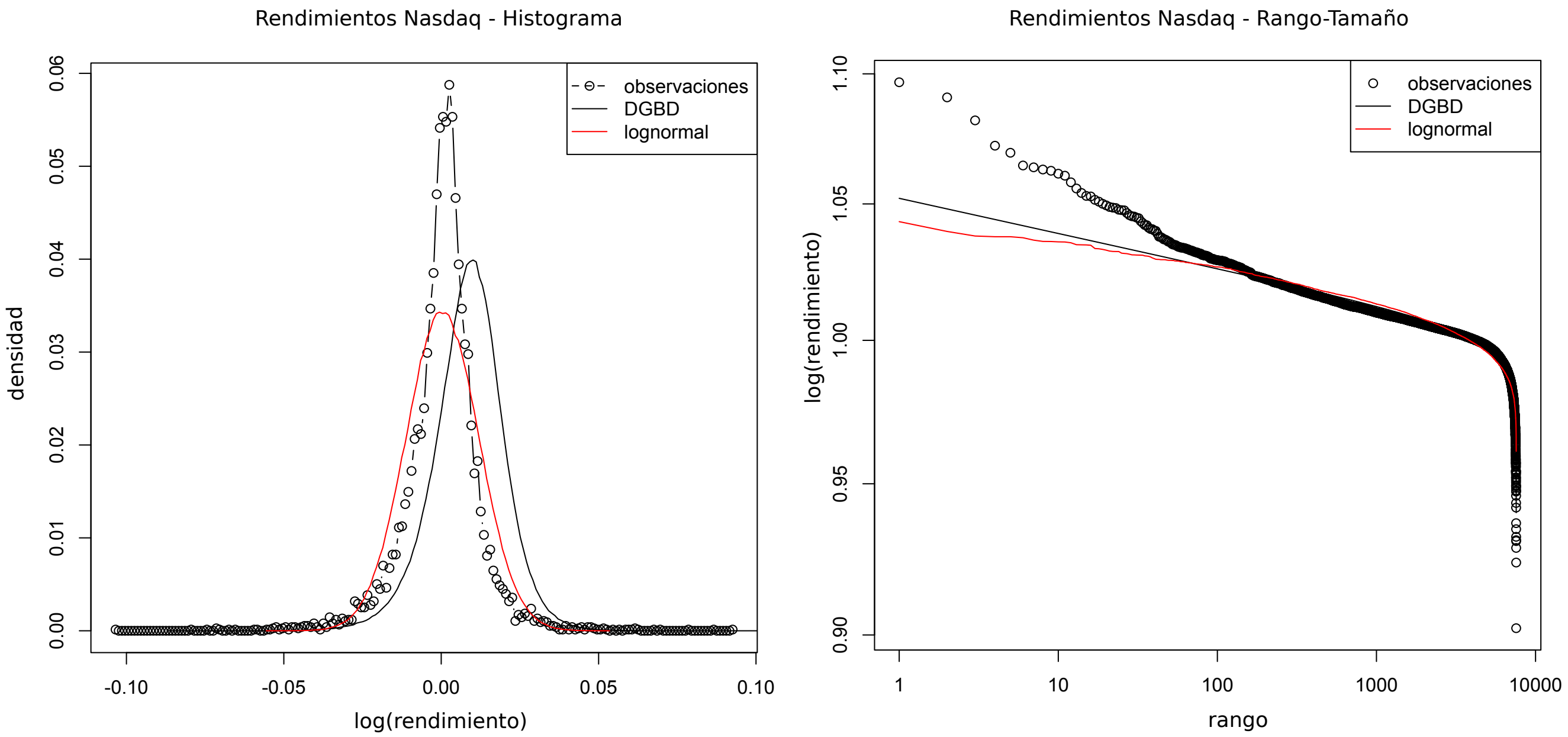

*Figura 6. Histograma y rango-tamaño en escala logarítmica para los rendimientos diarios, respecto al precio más alto, del índice Nasdaq 100 para un período de 30 años. En ambos casos mostramos los datos empíricos así como la distribución lognormal y la función DGBD ajustadas.*

La teoría de la probabilidad, a través de las leyes de grandes números y de los teoremas de límite central, nos permite modelar una enorme cantidad de fenómenos aleatorios y de hacer predicciones estadísticas sobre el comportamiento colectivo de sistemas que, aunque son estocásticos en principio, muestran patrones de comportamiento global perfectamente regulares y predecibles. En lo que respecta a los sistemas complejos, se ha observado en muchas ocasiones que las propiedades de interés siguen distribuciones de probabilidad de cola pesada.

Cuando un fenómeno sigue una distribución de cola pesada, sucede que estadísticas como el promedio y la varianza muestral no son informativas, pues no existe una escala característica para la ocurrencia del fenómeno. En este tipo de sistemas, los eventos extremos tienen una probabilidad de ocurrir relativamente alta; en muchas ocasiones, son los eventos extremos los que más nos interesa estudiar, comprender y eventualmente predecir, pues su impacto es mucho mayor que el de todos los eventos no extremos acumulados, tal como ocurre con las grandes caídas de los mercados o los terremotos de gran magnitud.

El modelo clásico dentro de la teoría de los sistemas complejos para describir fenómenos de cola pesada es la ley de potencias. Se conocen muchos mecanismos teóricos que explican su aparición y se ha reportado que una enorme cantidad de fenómenos es gobernada por este tipo de leyes; sin embargo, validarlas a través de datos empíricos es una tarea que dista de ser trivial, además de que existen motivos teóricos para pensar que muchos de estos sistemas sufren en la práctica de efectos de tamaño finito, por lo cual debe haber al menos dos dominios de descripción: escalas grandes y escalas pequeñas.

Existen varias alternativas para modelar esta clase de fenómenos. Aquí hemos expuesto una de ellas, que es relativamente nueva: la función DGBD. La evidencia estadística que hay en muchos ejemplos favorable hacia este modelo y los mecanismos teóricos que se están estudiando para explicar su aparición la convierten en una alternativa que, creemos, es digna de ser considerada y estudiada con mayor profundidad.

**Referencias**